\documentclass[12pt]{iopart}
\usepackage{graphicx}

\begin{document}

\title{The spectrum of a quantum potential}
\author{Paolo Amore\footnote{paolo@cgic.ucol.mx}}
\address{Facultad de Ciencias, Universidad de Colima, \\
Bernal D\'{\i}az del Castillo 340, Colima, Colima, M\'exico} 
\author{Jorge A. L\'opez}
\address{Physics Department, University of Texas at El Paso, \\
  El Paso, Texas, USA}

\begin{abstract}    
We apply a method recently devised by one of the authors to obtain precise analytical 
formulas for the spectrum of quartic and sextic anharmonic potential.
Due to its general features the method can be applied with minimal effort to a wide class
of quantum potentials thus allowing very promising applications.

\end{abstract}
\pacs{03.65.Ge,02.30.Mv,11.15.Bt,11.15.Tk}      
\maketitle


Recently one of the authors has developed a method which allows to obtain 
analytical approximations to a certain class of integrals with arbitrary 
precision~\cite{AS:04,AAFS:04}. It has been proven that the expansion proposed 
in \cite{AS:04} converges uniformly and that it yields excellent
results already to first order. Such method has been applied to a large 
class of problems in Classical Mechanics and General Relativity, allowing 
to obtain simple analytical formulas which work even in the non-perturbative
regime.

In this letter we deal with a further application of the method, showing that
it is possible to obtain accurate {\sl analytical} expressions for the spectrum 
of a quantum potential by using it in conjunction of the 
WKB expansion. Although our calculations are obtained for the anharmonic potential,   
the procedure that we suggest applies with minor changes to any potential. 

We will use \cite{BOW:77} as our main reference: in that paper Bender and collaborators 
show how to solve the one-dimensional eigenvalue problem for analytic potential
to all orders in the WKB approximation. They obtain a recursion relation for the WKB 
corrections and they are able to sum the WKB series in two special cases, recovering 
the exact energy eigenvalues.

However, in the general case of a potential $V(x)$ the calculation of the WKB correction to a
given order can be difficult and, depending on the potential itself, it is possible
that no analytical expression can be found for the WKB integrals. By applying the method of
\cite{AS:04} to the anharmonic oscillator to order $\hbar^4$ we show that one can obtain
a precise {\sl analytical} approximation to the {\sl entire spectrum} of the potential.
The precision of the approximation depends both on the order to which  the WKB expansion is considered
and on  the order to which our method is applied.

Let us briefly review the method by considering the integral ${\cal J}_1 = \int_{x_-}^{x_+} \ \sqrt{E-V(x)} \ dx$.
$x_\pm$ are the classical turning points of the potential $V(x)$ and $E = V(x_\pm)$ is the energy.
In the spirit of the Linear Delta Expansion~\cite{lde} we interpolate the potential $V(x)$ as
\begin{eqnarray}
V_\delta(x) = V_0(x) + \delta \ (V(x)-V_0(x)) \ ,
\label{EQN1}
\end{eqnarray}
where $V_0(x)$ is a potential introduced by hand, which depends on one or more arbitrary parameters\footnote{In the 
following we call $\lambda$ these parameters.}. 
$V_\delta(x)$ reduces to the full potential for $\delta=1$.
We want to perform an expansion described by eq.~(\ref{EQN1}) without moving the inversion points;
for this reason we ask that $x_\pm$ be the inversion points also of the potential $V_0(x)$. 
As a result, the energy $E_0$ that the particle would possess if it was moving only in the potential $V_0(x)$ will
be given by:
\begin{eqnarray}
E_0 = V_0(x_\pm) \ .
\end{eqnarray}

We therefore write the integral as
\begin{eqnarray}
{\cal J}_1 &=&   \int_{x_-}^{x_+} \ \sqrt{E_0-V_0(x)} \ \left[ 1 + \delta \Delta(x) \right]^{1/2} \ dx
\end{eqnarray}
where
\begin{eqnarray}
\Delta(x) \equiv \frac{E -E_0 - V(x) + V_0(x)}{E_0-V_0(x)}  .
\end{eqnarray}

We treat the term proportional to $\delta$ as a perturbation and perform an expansion in $\delta$.
Such expansion will converge uniformly when $|\Delta(x)| < 1$ in the 
region $x_- \leq x \leq x_+$~\cite{AS:04}. This condition selects a particular region in the parameter 
space, i.e. $\lambda$; however, maximal convergence is achieved when the Principle of Minimal 
Sensitivity (PMS)~\cite{Ste81} is used, i.e. when the condition $d{\cal J}_1/d\lambda = 0$ is enforced. 
Notice that if the potential $V_0(x)$ is chosen appropriately 
it will be possible to calculate analytically each term in the expansion.

We now come explicitly to the WKB integrals. We perform our analysis to order $\hbar^4$ and use the recursion 
formulas of \cite{BOW:77} to calculate the spectrum of the anharmonic oscillator. Following \cite{DT:00} we write 
the integral in a fashion where the singularities in the integrand are integrable. The price that we pay is that 
we have to introduce derivatives with respect to the energy.

We therefore write the WKB condition as
\begin{eqnarray}
\Lambda(E) = \frac{\pi \hbar}{\sqrt{2 m}} \ (n+1/2)
\label{EQN2}
\end{eqnarray}
where $\Lambda(E)$ to order $O(\hbar^6)$ is given by
\begin{eqnarray}
\Lambda(E) &\equiv& {\cal J}_{1}(E) -  \frac{\hbar^2}{48 \ m} \frac{d}{dE} {\cal J}_{2}(E) + 
\frac{\hbar^4}{11520 \ m^2} \frac{d^3}{dE^3}  {\cal J}_{3}(E) \ .
\label{EQN3}
\end{eqnarray}
We have defined:
\begin{eqnarray}
{\cal J}_{1}(E) &\equiv&  \int_{x_-}^{x_+} \sqrt{E-V(x)} dx \\
{\cal J}_{2}(E) &\equiv&  \int_{x_-}^{x_+} \frac{V''(x)}{\sqrt{E-V(x)}} dx \\
{\cal J}_{3}(E) &\equiv& \int_{x_-}^{x_+} \frac{7 \ V''(x)^2 - 5 \ V'(x) \ V'''(x)}{\sqrt{E-V(x)}} dx
\end{eqnarray}
where $x_{\pm}$ are the classical turning points. The spectrum of the potential $V(x)$ can be obtained by 
solving Eq.~(\ref{EQN3}).

So far we have used general formulas;  let us consider now the specific case of the anharmonic oscillator 
$V(x) = \frac{m \omega^2 x^2}{2} + \frac{\mu x^4}{4}$. 
We choose the interpolating potential $V_0(x) = \frac{1}{2} \left( m \omega^2 + \lambda^2 \right) \ x^2$. 
The method of \cite{AS:04} allows to calculate the integrals in Eq.~(\ref{EQN3}) quite simply. 
Indeed to order $\delta^{10}$ and using the value of $\lambda$ obtained by the PMS to first order one obtains:
\begin{eqnarray}
{\cal J}_{1}(\zeta) &=& \frac{m^{3/2} \ \pi \ \omega^3}{
\mu \ \zeta^{\frac{3}{2}} \ \left( 5 + 8 \ \zeta  \right)^{\frac{19}{2}}} \ 
\sum_{n=0}^{10} \ c_n^{(1)} \ \zeta^n
\label{EQN4a} \\
{\cal J}_{2}(\zeta) &=& \frac{{\sqrt{m}}\ \pi \ \omega \  
{\left( 3 + 2 \ \zeta  \right) }^{\frac{3}{2}}}{{\sqrt{\zeta }}\ 
{\left( 21 + 36\ \zeta  + 16\ {\zeta }^2 \right) }^{\frac{21}{2}}}  \ \sum_{n=0}^{20} \ c_n^{(2)} \ \zeta^n
\label{EQN4b} \\
{\cal J}_{3}(\zeta) &=& \frac{m^{\frac{3}{2}}\ \pi \ \omega^3\ 
    {\left( 99 + 48\ \zeta  + 56\ {\zeta }^2 \right) }^{\frac{3}{2}}}{{\sqrt{2}}\ {\zeta }^{\frac{3}{2}}\ 
    {\left( 363 + 564\ \zeta  + 360\ {\zeta }^2 + 
        224\ {\zeta }^3 \right) }^{\frac{21}{2}}} \ \sum_{n=0}^{30} \ c_n^{(3)} \ \zeta^n 
\label{EQN4c} 
\end{eqnarray}
where we have introduced the variable $\zeta \equiv \frac{m \ \omega^2}{\mu \ A^2}$ and $x_\pm= \pm A$ are the 
classical turning points. The coefficients $c_n^{(i)}$ are pure numbers, which we will not display here.

\begin{figure}
\begin{center}
\includegraphics[width=9cm]{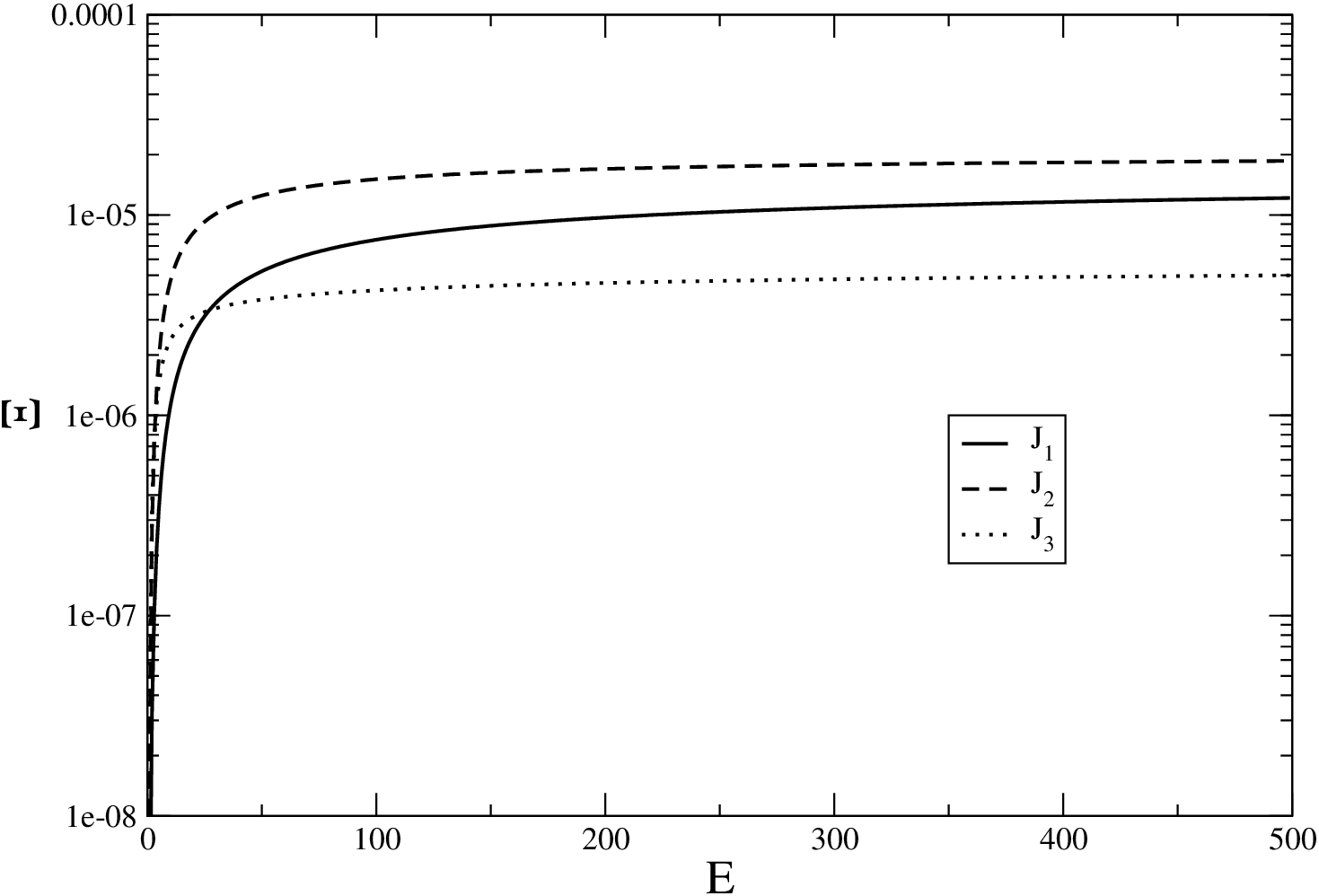}
\caption{Error over the integral ${\cal J}_i$ for $m = \omega = \hbar = \mu = 1$ as a function of the energy.}
\label{FIG1}
\end{center}
\end{figure}

In Fig.~\ref{FIG1} we plot the error defined as  
$\Xi_i = \left| \frac{{\cal J}_i^{(\delta)} - {\cal J}_i^{(exact)}}{{\cal J}_i^{(exact)}} \right| \ \times 100$ as
a function of the energy. We exploit the relation between the amplitude and the energy, i.e. $E = V(A)$, to express
$\zeta$ as a function of the energy:
\begin{eqnarray}
\zeta = \frac{m^2 \omega^4}{4 \mu E_n} \ \left[ 1 + \sqrt{1 + \frac{4  \mu  E_n}{m^2  \omega^4}} \right] \ .
\label{EQN5}
\end{eqnarray}
${{\cal J}_i^{(exact)}}$ are the exact integrals. We have assumed the values $m = \omega = \hbar = \mu = 1$. 

We observe that the error is smaller at lower energies and flattens around values of the order
of $10^{-5}$ at large energies. Such result can be further improved by applying our method to higher orders
in $\delta$: indeed it was proved in \cite{AS:04} that the method is convergent and therefore the precision of the calculation
can be easily improved. We stress that the behavior of this error is somewhat complementary to the one expected for the
WKB expansion, which is known to work better for the highly excited states. This is of course a nice feature, which 
allows to obtain a higher precision in our calculations.

Once Eqs.~(\ref{EQN4a}), (\ref{EQN4b}) and (\ref{EQN4c}) are substituted back in Eq.~(\ref{EQN3}) one obtains a 
rather complicated expression, which still needs to be inverted in order to give the energy as a function of 
the quantum number $n$. 
However this task is easily accomplished by Taylor expanding Eq.~(\ref{EQN3}) around  $\zeta = 0$ 
(which corresponds to the higher part of the spectrum) and by then using eq.~(\ref{EQN5}).

We obtain the approximate {\sl analytical} formula for the spectrum of the quartic anharmonic oscillator:
\begin{eqnarray}
E_n &\approx& e_1 \ \left(n+\frac{1}{2}\right)^{\frac{4}{3}} + e_2 \ \left(n+\frac{1}{2}\right)^{\frac{2}{3}} + 
e_3 + \frac{e_4}{\left( n+\frac{1}{2}\right)^{2/3}} + \dots 
\label{EQN6}
\end{eqnarray}
where the first few coefficients $e_i$ are given by
\begin{eqnarray}
e_1 &=& \frac{8192000000000000\,{\left( \frac{2}{371496873} \right) }^{\frac{1}{3}}\,5^{2/3}}{48413843986233} \ 
{\left( \frac{{\mu}\ {{\hbar}}^4}{m^2} \right) }^{\frac{1}{3}} \approx 0.867146 \ 
{\left( \frac{{\mu}\ {{\hbar}}^4}{m^2} \right) }^{\frac{1}{3}} \\
e_2 &=& \frac{223553096351603200000\,{\left( \frac{2}{371496873} \right) }^{2/3} \ 
5^{1/3}}{2759589107215281} \ \left( \frac{\hbar \ m}{\sqrt{\mu}} \right)^{2/3} \ \omega^2 \nonumber \\
&\approx& 0.42551  \left( \frac{\hbar \ m}{\sqrt{\mu}} \right)^{2/3} \ \omega^2 \\
e_3 &=& - \frac{22736856054709769417272276}{486959893945285848925248675} \ \frac{m^2 \ \omega^4}{\mu} \approx 
-0.0466914 \ \frac{m^2 \ \omega^4}{\mu} \\
e_4 &=& \frac{9339788533820875\,5^{\frac{1}{6}} }{131900122669056\,2^{\frac{2}{3}}\,
     {\sqrt{7}}\,{371496873}^{\frac{1}{3}}} \ \left( \frac{\mu \hbar^4}{m^2} \right)^{1/3}
\nonumber \\
&+& \frac{8271957929254630511939617633398178757}{999241702375726561994610281100000000\,2^{\frac{2}{3}}\,
     {1857484365}^{\frac{1}{3}}}  \ \left( \frac{m^{10} \omega^6}{\mu^5 \hbar^2}\right)^{1/3} \nonumber \\
&\approx& 0.030669 \left( \frac{\mu \hbar^4}{m^2} \right)^{1/3} + 0.00424238 \  
\left( \frac{m^{10} \omega^6}{\mu^5 \hbar^2}\right)^{1/3} \ .
\end{eqnarray}

Notice that altough the coefficients $e_i$ of higher order could be calculated as well, it is not necessary to do so, 
since they would only provide corrections that would be dominated by the error that we are making evaluating  the 
integrals ${\cal J}_i$. 

\begin{figure}
\begin{center}
\includegraphics[width=9cm]{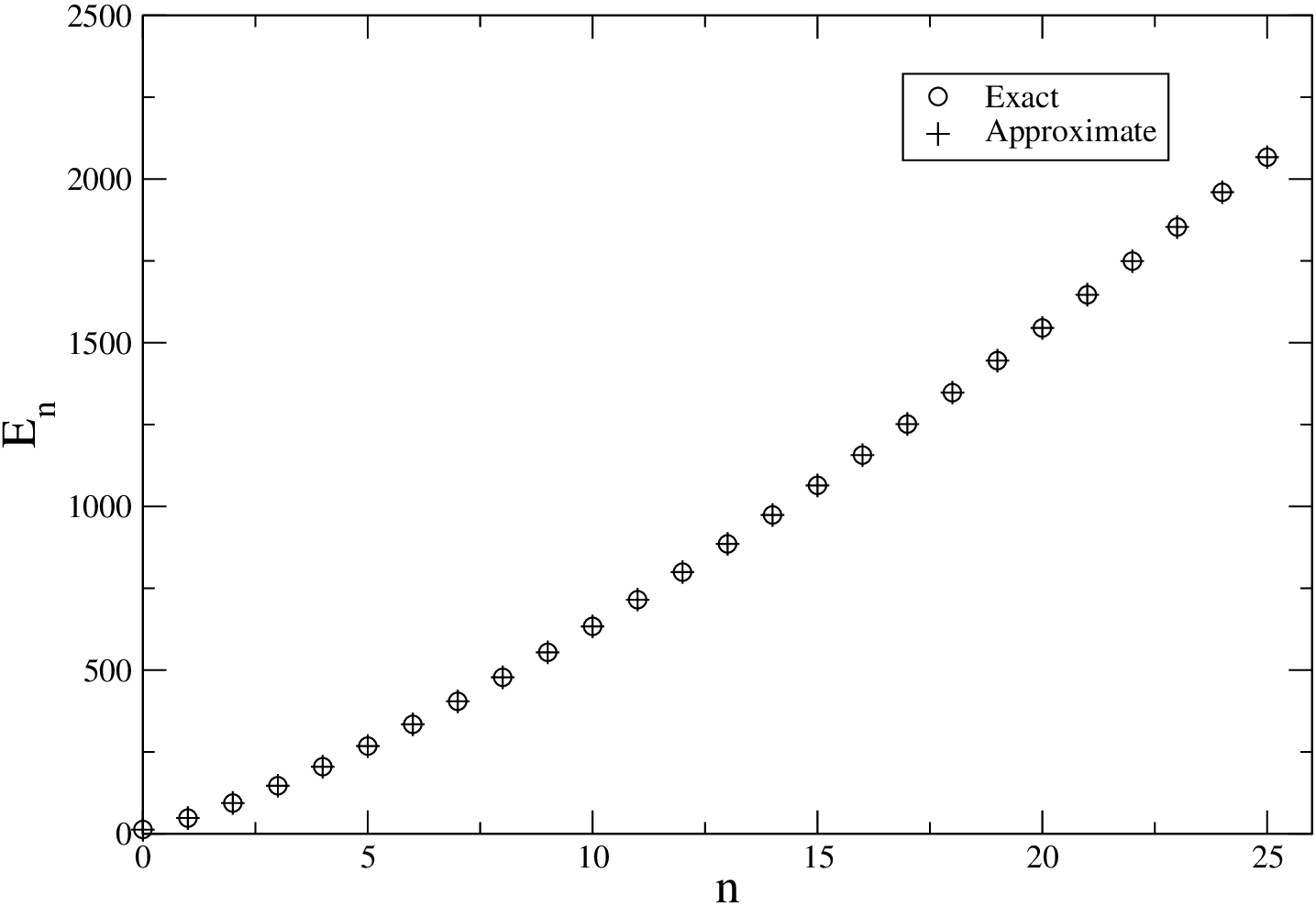}
\caption{Comparison between the numerical eigenvalues of \cite{mei97} and the results obtained using eq.~(\ref{EQN6}). 
We assume  $\hbar = 1$, $m= 1/2$, $\omega= 2$ and $\mu = 8000$.}
\label{FIG2}
\end{center}
\end{figure}

In  Fig.~\ref{FIG2} we plot the energy eigenvalues of the last column of Table III of \cite{mei97}, corresponding to 
 $\hbar = 1$, $m= 1/2$, $\omega= 2$ and $\mu = 8000$ and compare them with the approximate results obtained 
by using eq.~(\ref{EQN6}). It is clear that our formula, although quite simple, works excellently.

\begin{figure}
\begin{center}
\includegraphics[width=9cm]{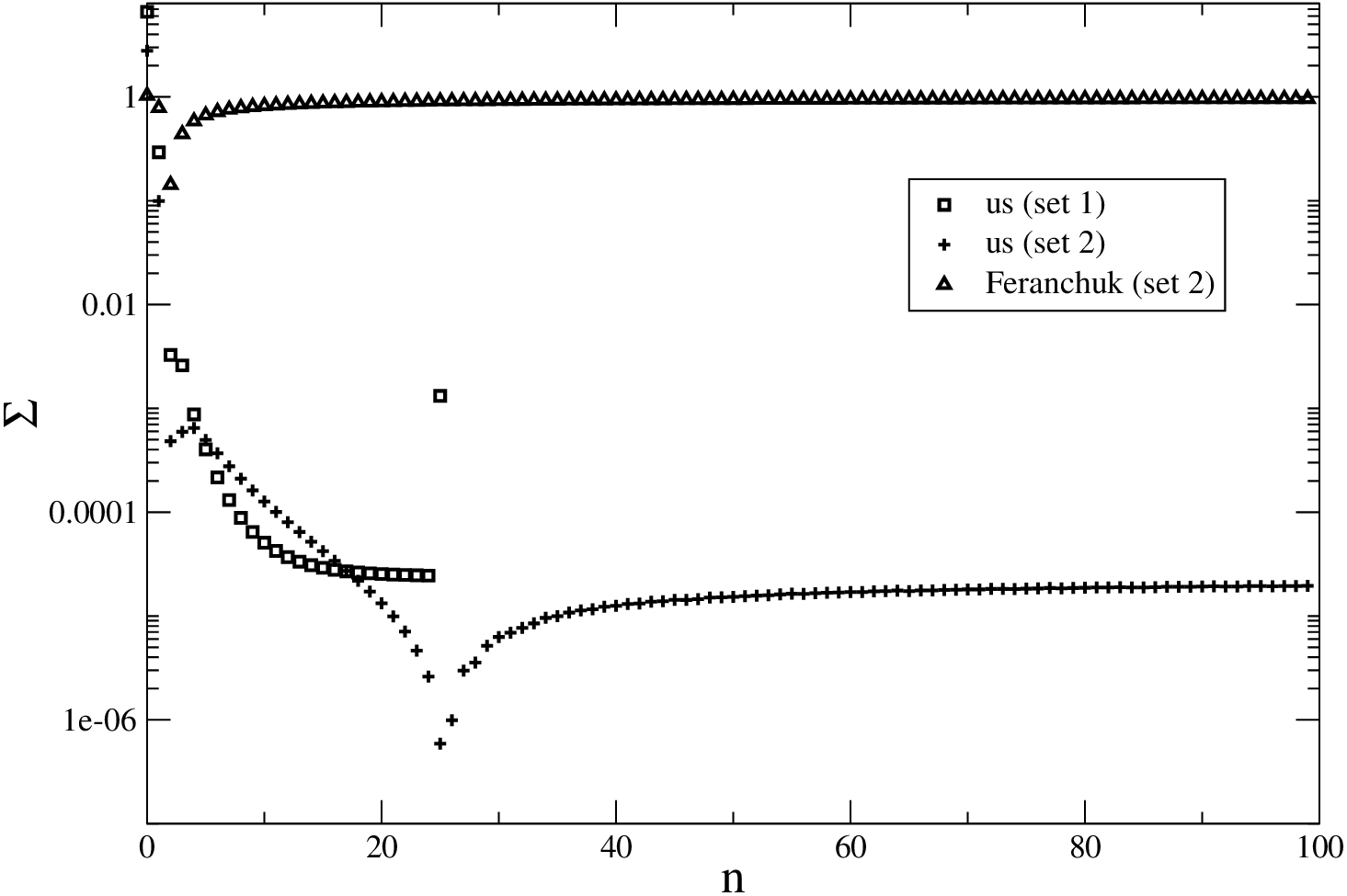}
\caption{Error over the energy of the anharmonic oscillator with  $\hbar = 1$, $m= 1/2$, $\omega= 2$ and $\mu = 8000$ (set 1) and
with $\hbar = m= \omega= 1$ and $\mu = 4$ (set 2). The boxes and the pluses have been obtained with our method, the triangles correspond to the 
error calculated using the analytical formula of \cite{Fer95}.}
\label{FIG3}
\end{center}
\end{figure}

In Fig.~\ref{FIG3} we display the error over the energy defined as 
$\Sigma \equiv \left| \frac{E_n^{(approx)}-E_n^{(exact)}}{E_n^{(exact)}} \times 100\right|$ as a function of the quantum 
number $n$. The boxes have been obtained using our formula eq.~(\ref{EQN6}) and assuming $\hbar = 1$, $m= 1/2$, $\omega= 2$ and $\mu = 8000$. 
In this case $E_n^{(exact)}$ are the energies of the anharmonic oscillator calculated with high precision in last column of Table III of 
\cite{mei97}. The jump corresponding to $n=25$ is due to the low precision of the last value of Table III of \cite{mei97}. 
The pluses and the triangles have been obtained using  our formula eq.~(\ref{EQN6}) (pluses) and eq.~(1.34) of \cite{Fer95} (triangles)
and assuming $\hbar = m= \omega= 1$ and $\mu = 4$. In this case $E_n^{(exact)}$ are the energies of the anharmonic oscillator numerically 
calculated through a fortran code. We can easily appreciate that our formula provides an approximation which is several order of magnitude better
than the one of eq.~(1.34) of \cite{Fer95}.

As a further check on the soundness of our approximation we consider the quantum zeta function for the quartic anharmonic 
oscillator~\cite{Vor80,CR96}:
\begin{eqnarray}
Z(s) = \sum_{n=0}^\infty \frac{1}{E_n^s}
\end{eqnarray}

The exact value of $Z(1)$ corresponding to setting $\omega=0$ and $\mu=4$ is known to be
\begin{eqnarray}
Z(1) = \frac{3^{2/3}}{8 \pi^2} \Gamma^5\left(\frac{1}{3}\right) \approx 3.63500364488 \ .
\end{eqnarray}

We have estimated $Z(1)$ by using the first few numerical eigenvalues and then using for the remaining ones 
the analytical formula corresponding to the solution of  eq.~(\ref{EQN6}). Using our formula eq.~(\ref{EQN6})  only
taking the first $4$ eigenvalues numerically we have obtained the value
\begin{eqnarray}
Z_{4}(1) = 3.635002
\end{eqnarray}
which can be compared with the value 
\begin{eqnarray}
Z_{140}(1) = 3.6350017
\end{eqnarray}
obtained in \cite{CR96} by using a much larger set of numerical eigenvalues.

\begin{figure}
\begin{center}
\includegraphics[width=9cm]{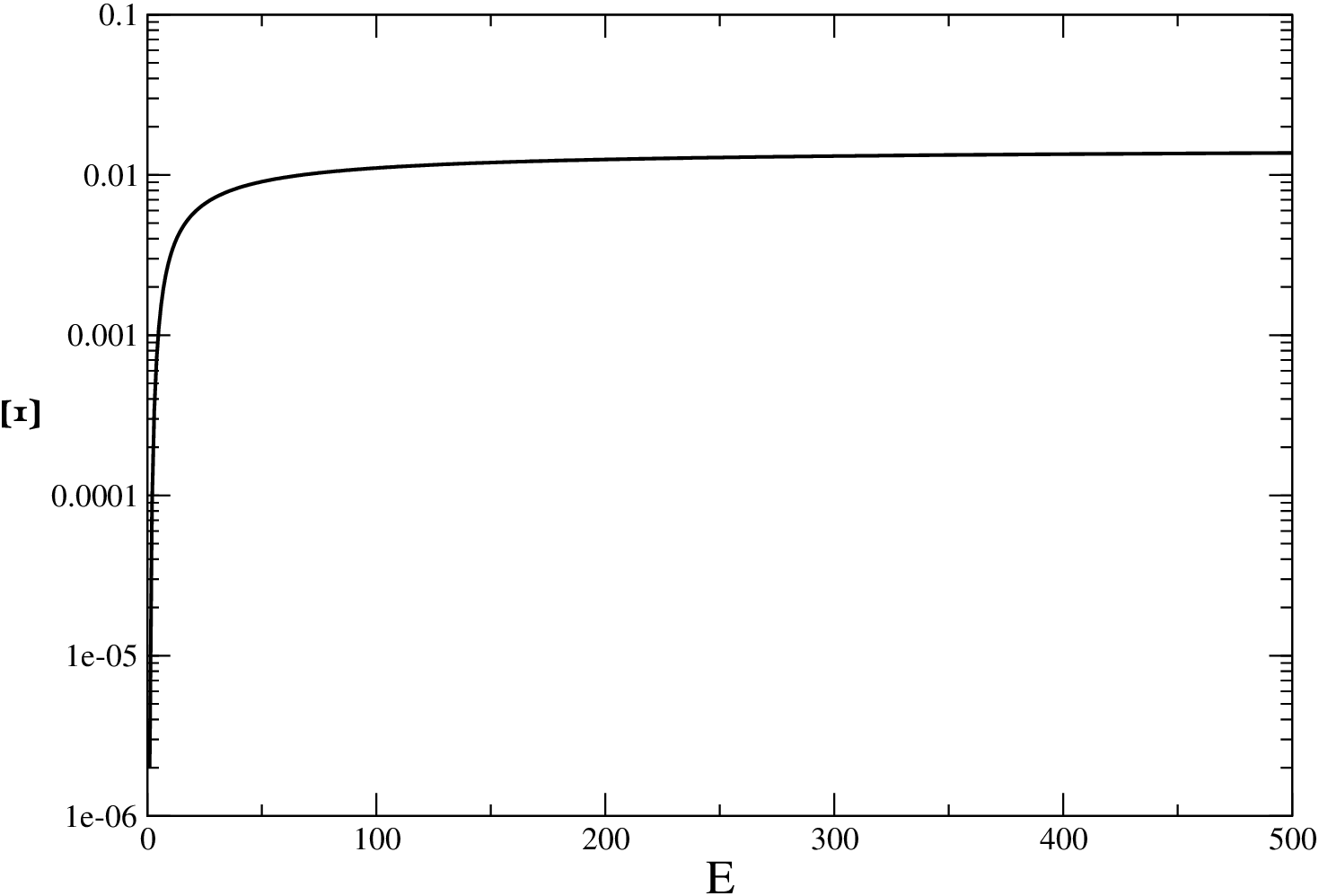}
\caption{Error over the integral $J_1$ for $\hbar = m = \omega =\rho =1$.}
\label{FIG4}
\end{center}
\end{figure}

We have also applied our method to the case of a anharmonic potential of the form
$V(x) = \omega x^2/2 + \rho x^6/6$.
In this case we have used the WKB expansion to leading order and we have approximated with our method the  
integral ${\cal J}_{1}$ corresponding to this potential. In Fig.~\ref{FIG4} we display the error 
$\Xi = \left| \frac{{\cal J}_1^{(\delta)} - {\cal J}_1^{(exact)}}{{\cal J}_1^{(exact)}} \right| \ \times 100$ as
a function of the energy. ${\cal J}_1^{(exact)}$ is the exact value of the integral calculated numerically, whereas
${\cal J}_1^{(\delta)}$ is the analytical approximant obtained with our method to order $10$. 
Although  the error increases with the energy, we notice that it is always smaller than $0.01 \%$. As in the
case of the quartic oscillator, our approximation works better in the region of low energy, 
i.e. opposite to the WKB method. 

By using the approximation for the integral ${\cal J}_{1}$ obtained with our method we have derived a very simple
formula for the spectrum of the sextic oscillator to leading order in the WKB expansion:
\begin{eqnarray}
E_n &=& \frac{1}{\left( \alpha_1 \left(n+\frac{1}{2}\right) + \alpha_2 - \sqrt{\beta_1  \left(n+\frac{1}{2}\right)^2 +
\beta_2  \left(n+\frac{1}{2}\right) + \beta_3} \right)^{3/2}}
\label{EQN7}
\end{eqnarray}
where
\begin{eqnarray}
\alpha_1 &=& \frac{20887573200705159168\,6^{\frac{1}{6}}\,{\sqrt{11}}\, {\hbar}\,
    \,{\rho}^{\frac{5}{6}}}{5057378805833647189\, m^{\frac{5}{2}}\,\omega^4} \approx 18.46505979 \ 
\frac{\hbar \rho^{5/6}}{m^{5/2}} \nonumber \\
\alpha_2 &=& \frac{8129630071150442107\,6^{\frac{2}{3}}\,{\rho}^{\frac{1}{3}}}
  {5057378805833647189\,m\,\omega^2} \approx 5.307778611 \ \frac{\rho^{1/3}}{m \omega^2} \nonumber \\
\beta_1 &=& \frac{4799197856362980042268993891174358974464\,6^{\frac{1}{3}}\, 
{{\hbar}}^2\,{\rho}^{\frac{5}{3}}}{255770803856953672746878259\ 25727601721\,m^5\,\omega^8} \nonumber \\ 
&\approx& 340.9584332 \ \frac{\hbar^2 \rho^{5/3}}{m^5 \omega^8} \nonumber \\
\beta_2 &=& \frac{339616486411617501798889917918008573952\,6^{\frac{5}{6}}\,
    {\sqrt{11}}\, {\hbar}\,{\rho}^{\frac{7}{6}}}{255770803856953672\
74687825925727601721\,m^{\frac{7}{2}}\, \omega^6} \nonumber \\
&\approx& 196.0168989 \frac{\hbar \rho^{7/6}}{m^{7/2} \omega^6}\nonumber \\
\beta_3 &=& -\frac{177921136833163170599693588290032640752\,6^{\frac{1}{3}}\ 
{\rho}^{\frac{2}{3}}}{25577080385695367274687825925727601721\,m^2\,\omega^4} \nonumber \\
&\approx& -12.64038572 \frac{\rho^{2/3}}{m^2 \omega^4} \nonumber  \ .
\end{eqnarray}

\begin{figure}
\begin{center}
\includegraphics[width=9cm]{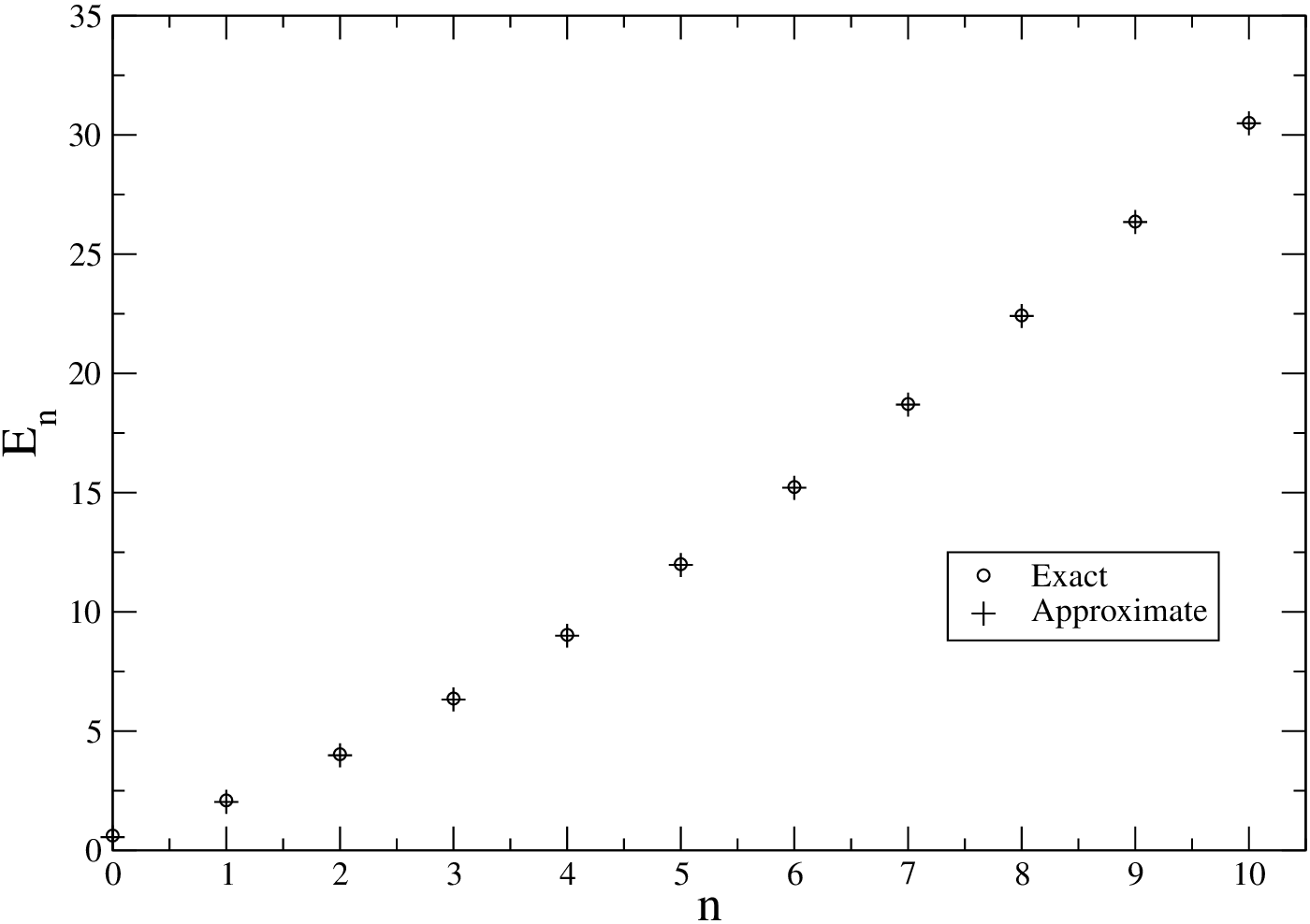}
\caption{Lower part of the spectrum of the sextic anharmonic oscillator. The circles are the exact 
eigenvalues calculated numerically and the pluses are the results of eq.~(\ref{EQN7}). 
We assume $\hbar = m = \omega =\rho =1$.}
\label{FIG5}
\end{center}
\end{figure}

\begin{figure}
\begin{center}
\includegraphics[width=9cm]{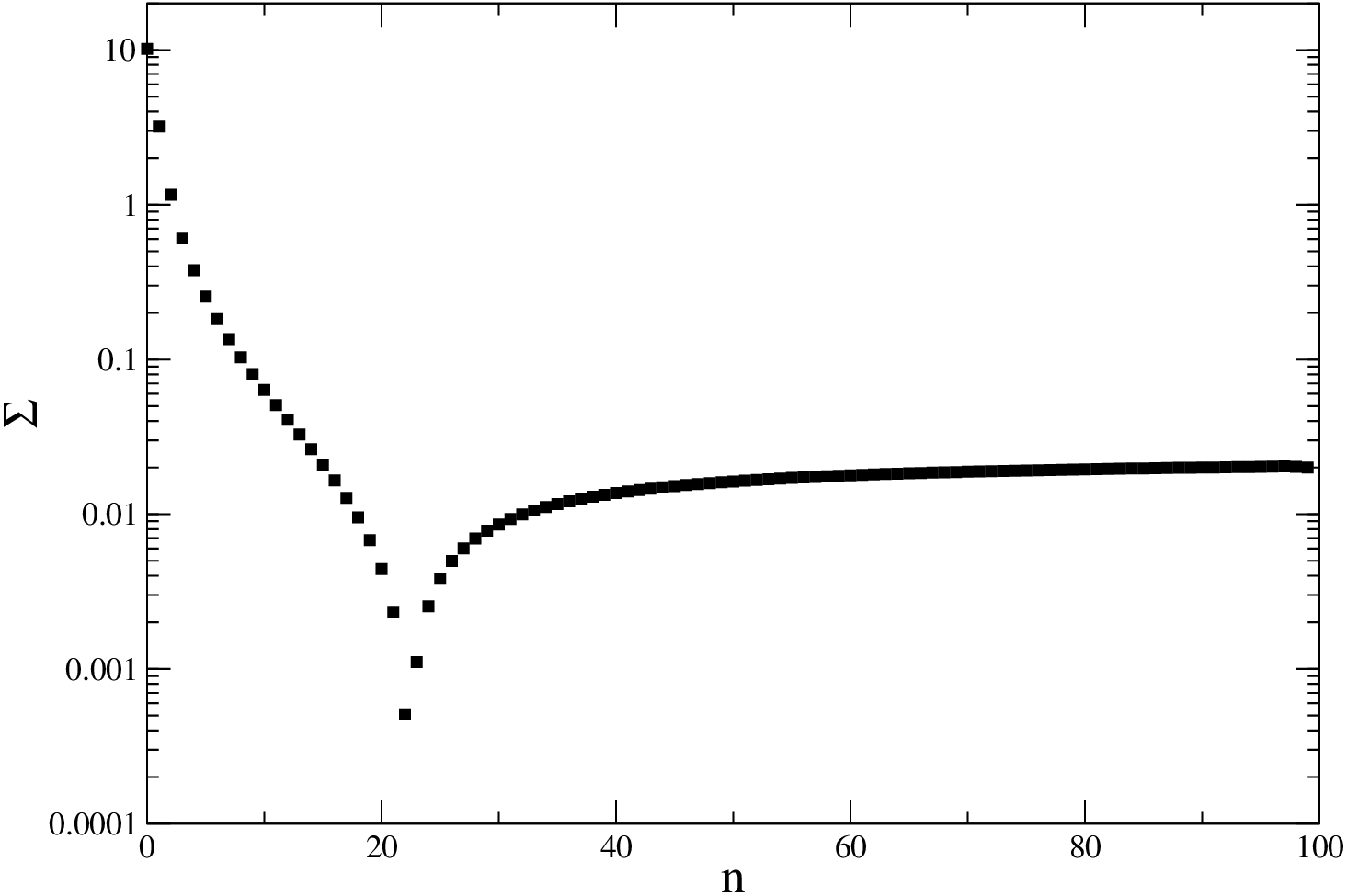}
\caption{Error over the energy obtained using eq.~(\ref{EQN7}) and assuming $\hbar = m = \omega =\rho =1$.}
\label{FIG6}
\end{center}
\end{figure}

In Fig.~\ref{FIG5} we compare the lower part of the spectrum obtained using eq.~(\ref{EQN7}) 
with the exact eigenvalues, obtained numerically with a fortran code. Once again the agreement between the 
two is impressive. 
Fig.~(\ref{FIG6}) displays the error (in $\%$) over the energy obtained using eq.~(\ref{EQN7}). 
The error is of the order of a few percents for low values of $n$, but drops quite quickly, reaching a plateau 
of about $0.01 \%$ at large $n$. 
The presence of the plateau is a consequence of the error that we are making 
in the evaluation of ${\cal J}_1$, as already discussed when we considered Fig.~\ref{FIG4}. 
Since the method that we are using is convergent\cite{AS:04,AAFS:04}, by working to higher orders one can arbitrarily 
lower the position of such plateau, to obtain the highly excited part of the energy spectrum with the desired accuracy.

In conclusion we have shown that the method of \cite{AS:04,AAFS:04} can be easily applied to calculate fully analytical 
expressions for the energy spectrum of quantum potentials within the WKB approximation. By applying our method
to a sufficiently large order in $\delta$ we are able to estimate {\sl analytically} the WKB integrals with high
precision, given the convergent nature of the expansion that we are performing. The application of 
this method to more general potentials in $1$ and $3$ dimensions is currently underway.

\verb''\ack
The authors acknowledge useful conversations with Dr.~Alfredo Aranda.
P.A. acknowledges support of Conacyt grant no. C01-40633/A-1. J.A.L. thanks the Universidad de Colima for the warm
hospitality.

\verb''\section*{References}
\verb''

\end{document}